\documentclass[11pt,letterpaper]{article}
\pdfoutput=1
\usepackage{jheppub}
\usepackage{graphicx}
\input{epsf}
\usepackage{epsfig}
\usepackage{breqn}
\usepackage{epstopdf}
\usepackage{subfigure}
\usepackage{amsmath}
\usepackage{pgfplots}
\newcommand{\be}{\begin{equation}}
\newcommand{\ee}{\end{equation}}

\DeclareMathOperator{\Tr}{Tr}

\title{Higher-curvature corrections to holographic entanglement entropy in geometries with hyperscaling violation}
\author{Pablo Bueno}
\author{and Pedro F. Ram\'irez}
\affiliation{Instituto de F\'isica Te\'orica UAM/CSIC,\\C/ Nicol\'as Cabrera, 13-15, C.U. Cantoblanco, 28049 Madrid, Spain}\vspace{0.2cm}
\emailAdd{p.bueno@csic.es}
\emailAdd{p.f.ramirez@csic.es}
\abstract{We study the effects of including higher-curvature corrections to the Einstein gravity bulk action on the holographic entanglement entropy (HEE) expression for geometries with hyperscaling violation (hvLf). For $\theta< 0$ we show that one single new divergence arises for general curvature-squared gravities, which allows us to conjecture the general expression of HEE for any higher-order gravity action. For $0<\theta<d$, we assume the hvLf geometry to arise above some intermediate scale $r_F$, becoming AdS in the UV and perform a similar analysis for $R^n$ gravities. For negative values of $\theta$ we find that new logarithmic contributions show up in the HEE formula for any $n$th-order gravity when $\theta=d(d-1)/(d-2(n-1))$ and $d<2(n-1)$. In the range $0\leq \theta<d$ we do not find additional logarithmic contributions appearing at any order except for $n=1$, which corresponds to the famous case $\theta=d-1$ encountered in Einstein gravity. }

\preprint{IFT-UAM/CSIC-14-088
}
\arxivnumber{1408.6380}

\begin{document}
\maketitle

\section{Introduction}\label{Introduction}

The gauge/gravity duality \cite{Maldacena:1997re,Witten:1998qj,Gubser:1998bc} has proven to be an outstandingly successful and fruitful framework for probing the physics of strongly coupled field theories. The paradigmatic AdS/CFT correspondence, which established the physical equivalence between $d=4$, $\mathcal{N}=4$ Super-Yang-Mills and type-IIB String Theory on AdS$_5\times$S$^5$ \cite{Maldacena:1997re} has been extended over the years in a variety of ways in the hope of accounting for the physics of more realistic quantum field theories, such as QCD and condensed matter systems (see, e.g., \cite{Aharony:1999ti,Polchinski:2010hw,CasalderreySolana:2011us,Hartnoll:2009sz} for reviews on these subjects).

One such extension consists of considering systems in which, albeit scaling symmetry is respected, space and time do not scale in the same way, so conformal (and Lorentz) invariance is broken. This is the case of the so-called Lifshitz fixed points, characterized by a \emph{dynamical critical exponent} $z$, which determines the anisotropic scaling in the time direction $t$
\be
t\rightarrow \lambda^z t\, ,\, x_i\rightarrow \lambda x_i\, ,\,  i=1,...,d\, ,
\ee
being $x_i$ the $d$ spatial dimensions of the $(d+1)$-spacetime in which the field theory under consideration is defined. The class of $(d+2)$-dimensional dual spacetime geometries with the appropriate symmetries can be written, in some coordinate system, as \cite{Koroteev:2007yp,Kachru:2008yh,Taylor:2008tg}
\be \label{Lifshitz}
ds^2=-\frac{L^2}{r^{2z}}dt^2+\frac{L^2}{r^2}\left[dr^2+d\vec{x}^2_{(d)} \right]\, ,
\ee
which reduces to AdS$_{d+2}$ in the Poincar\'e patch for $z=1$. Embedding solutions of this kind (and others which asymptote to them) into gravity and String Theory models and studying their properties in the holographic framework has been subject of study in numerous previous works (see, e.g. \cite{Hartnoll:2009ns,Balasubramanian:2010uk,Donos:2010tu,Gregory:2010gx,Chemissany:2011mb,Cassani:2011sv,Halmagyi:2011xh,Gouteraux:2011ce,Bertoldi:2010ca}), and remains an active area of research.\\ \\
\noindent
$\circ$ \emph{Lifshitz metrics with hyperscaling violation.\\}
A further generalization can be achieved by considering the following family of spacetime metrics \cite{Charmousis:2010zz}
\be \label{hvLf}
ds^2=L^2r^{\frac{2(\theta-d)}{d}}\left[-r^{-2(z-1)}dt^2+dr^2+d\vec{x}^2_{(d)} \right]\, .
\ee
These geometries (which are conformally Lifshitz) include, in addition to $z$, another exponent, customarily named $\theta$, and are characterized by the following transformation rules under rescalings of the coordinates
\be \label{scal}
t\rightarrow \lambda^z t\, ,\, x_i\rightarrow \lambda x_i\, ,\, r\rightarrow \lambda r\, ,\, ds^2\rightarrow \lambda^{\frac{2\theta}{d}}ds^2\, .
\ee
A system whose thermal entropy scales as $S_{\text{th.}}\sim T^{d}$ is said to possess a hyperscaling behaviour. When the dynamical exponent is present, this scaling gets modified to $S_{\text{th.}}\sim T^{\frac{d}{z}}$.
It can be seen that in field theories with the kind of scaling defined by (\ref{scal}), thermal entropy scales in turn as $S_{\text{th.}}\sim T^{\frac{d-\theta}{z}}$ \cite{Gouteraux:2011ce,Huijse:2011ef}, and so, from the thermodynamic point of view, $d-\theta$ acts as the effective number of space-like dimensions of the system \cite{Huijse:2011ef}. The fact that $S_{\text{th.}}$ does not scale with its naive power of the temperature corresponds therefore to a violation of the hyperscaling behaviour \cite{Huijse:2011ef,Fisher:1986zz} (the hyperscaling case being obviously $\theta=0$)\footnote{From the holographic perspective, this would correspond to the entropy of a black brane whose spacetime metric asymptotes to one of these solutions \cite{Iizuka:2011hg}.}, and the above class of metrics has been consequently named \emph{hyperscaling-violating Lifshitz metrics} (hvLf in short). Although the $r^{\frac{2\theta}{d}}$ factor spoils dimensional analysis in (\ref{hvLf}), this can be easily restored by including an additional scale $r_F$: $r^{\frac{2\theta}{d}}\rightarrow (r/r_F)^{\frac{2\theta}{d}}$, which we will often fix to $1$ henceforth.

In order to have a clear interpretation of a constant $r$ slice (with $r\rightarrow 0$) of the geometry defined by (\ref{hvLf}) as the boundary of the metric, we require $\theta<d$ from now on\footnote{The formulation of the holographic dictionary for hvLf geometries has been addressed in \cite{Chemissany:2014xpa,Chemissany:2014xsa}.}. From a different perspective, $\theta>d$ would correspond to a negative effective number of spatial dimensions according to the arguments previously explained. Also, when $\theta>0$, hvLf metrics suffer from a curvature UV-singularity in the Einstein frame: indeed, the Kretschmann invariant scales as $R_{\mu\nu\rho\sigma}R^{\mu\nu\rho\sigma}\sim r^{-4\theta/d}$. In appearance, this means that hvLf metrics with $\theta< 0$ are completely reliable in the UV, whereas those with $0<\theta<d$ need to be completed asymptotically, something which is usually performed through the assumption that spacetime is described by (\ref{hvLf}) only above some scale $r_F$, but asymptotes to some well-behaved solution, such as AdS$_{d+2}$, as $r<<r_F$. As explained in \cite{Chemissany:2014xsa}, this statement is imprecise. The authors argue that hvLf geometries with $\theta\neq 0$ typically require a UV-divergent (linear) dilaton, which allows one to tune the curvature singularity (appearing in the cases in which $0<\theta<d$) by changing to an appropriate Weyl frame, and completely absorb it in such scalar field. The linear running character of the dilaton is a characteristic feature of general hvLf backgrounds (with $\theta \neq 0$) so one needs to be careful when interpreting the UV physics from the field theory perspective not only for $\theta>0$, but also for $\theta<0$\footnote{We thank Robert C. Myers and Ioannis Papadimitriou for their comments and explanations about this point}. We will come back to this in the discussion section.

hvLf and asymptotically hvLf solutions have been extensively (and intensively) studied in the context of holography in e.g. \cite{Huijse:2011ef,Dong:2012se,Ogawa:2011bz,Shaghoulian:2011aa,Alishahiha:2014cwa,Alishahiha:2014jxa}. The gravity models in which solutions of this kind have been found and studied include for example Einstein-Maxwell-Dilaton (EMD) \cite{Cadoni:2009xm,Charmousis:2010zz,Perlmutter:2010qu,Iizuka:2011hg,Alishahiha:2012qu,Gath:2012pg,Kim:2012pd,Fonda:2014ula,Edalati:2013tma,Edalati:2012tc,Cremonini:2012ir}, Supergravity and String Theory \cite{Gubser:2009qt,Dong:2012se,Perlmutter:2012he,Narayan:2012hk,Ammon:2012je,Bueno:2012sd,Bueno:2012vx} and EMD plus curvature-squared terms \cite{O'Keeffe:2013nha,Ghodrati:2014spa,Knodel:2013fua}. The motivation for including higher-curvature terms in the gravitational action is in general motivated from the fact that these would correspond to $1/\sqrt{\lambda}$ corrections from the dual field theory perspective, allowing us to move slightly away from the infinitely coupled regime. In the particular case of Lifshitz and hvLf geometries, there are other reasons to include such corrections, such as changing the $(\theta,z)$ parameter space allowed by the null energy conditions (NEC) or curing the characteristic infrared (IR) divergent behaviour of the dilaton \cite{Huijse:2011ef} appearing in EMD theories (see \cite{O'Keeffe:2013nha} for details on these issues).\\\\
\noindent
\emph{$\circ$ Entanglement entropy in quantum field theories and the area law.\\}\noindent
There are several ways in which holography allows us to study the properties of the dual quantum field theories (QFTs). A prominent example is the computation of \emph{entanglement entropy} (EE), which will be the subject of this paper.

Entanglement entropy has indeed become an essential tool in fields as diverse as condensed matter \cite{2003quant.ph..4098L,2005PhRvL..94f0503P,2008RvMP...80..517A,Grover:2012sp}, quantum information \cite{Nielsen,2008PhRvL.100g0502W}, String Theory and quantum gravity \cite{Srednicki:1993im,Bombelli:1986rw,Callan:1994py,Bianchi:2012ev,Ryu:2006bv,Ryu:2006ef,Maldacena:2013xja,Nishioka:2006gr,Lewkowycz:2012mw,VanRaamsdonk:2009ar,Myers:2013lva,Headrick:2014eia}, and QFT \cite{Casini:2009sr,Calabrese:2004eu,Klebanov:2007ws,Kitaev:2005dm,Casini:2004bw,Shiba:2013jja,Casini:2013rba}.

For a particular QFT, given a spatial region $A$, EE is defined as: $S=-\Tr \left[ \rho_A \log \rho_A \right]$, being $\rho_A$ the reduced density matrix obtained by integrating out the degrees of freedom in the complement $\bar{A}$ (in this case, the entanglement entropy is also referred to as \emph{geometric} entropy, given that the Hilbert space separation is performed through the (artificial) geometric division of the spatial slice into two regions). The ultraviolet (UV) behaviour of the EE for general $(d+1)$-dimensional QFTs is expected to be \cite{Casini:2009sr}:
\be \label{see}
S=\frac{k_{d-1}}{\delta^{d-1}}+...+\frac{k_{1}}{\delta}+k_0\log{\frac{l}{\delta}}+S_0\, ,
\ee
where $\delta$ is a short distance cutoff, $S_0$, $k_0$ and $k_i$ constants, and $l$ is a characteristic length of $A$. The coefficient of the leading term is proportional to the \emph{area} of the boundary of $A$ ($k_{d-1}\sim l^{d-1}$), a behaviour which is usually argued to be caused by the entanglement between degrees of freedom living at both sides of $\partial A$. This is the so-called \emph{area law} \cite{Srednicki:1993im,Bombelli:1986rw} of entanglement entropy. When the leading term in EE depends on the characteristic length of $A$ in a different fashion, we speak about a violation of this law. One such kind of violation occurs when the leading contribution to $S$ contains a factor which scales logarithmically with the characteristic length of $A$ (see below). Another example of this happens when the leading term scales with a power of $l$ different from the dimension of $\partial A$ (see, e.g. \cite{2014arXiv1408.1657M}).

 An interesting point to notice is the fact that $k_0$ is universal in the following sense: if we shift $\delta\rightarrow  \delta\epsilon$, the coefficients $k_i$ are shifted by $k_i\rightarrow k_i \epsilon^{-i}$, whereas $k_0$ remains the same by virtue of the properties of the logarithm (the shift is absorbed in $S_0$). As a consequence, $k_0$ is independent of the regularization prescription (and usually related to the central charge of the underlying QFT in the case of CFTs). 

As we have said, although the area law turns out to hold for a vast range of systems, it is well-known that this is not always the case. A paradigmatic example is given by 2D CFTs, where EE scales logarithmically with the length of $A$, $l$, and $k_0$ turns out to be proportional to the central charge of the theory \cite{Holzhey:1994we,Calabrese:2004eu}
\be
S=\frac{c}{3}\log \frac{l}{\delta}\, .
\ee
In higher dimensional theories, violations of the area law appear in QFTs with Fermi surfaces \cite{Wolf:2006zzb,arxiv-0908.1724,Metlitski:2010pd}. In such cases, $S$ acquires a logarithmic dependence on the characteristic length of $A$
\be \label{S0fermi}
S\sim \left(l k_F\right)^{(d-1)} \log(l k_F)\, ,
\ee
being $k_F$ the Fermi momentum\footnote{Such behaviour comes from the effective 2D CFT which governs the physics of modes at the Fermi surface \cite{arxiv-0908.1724,Shaghoulian:2011aa}}, and the area law is violated. It has been argued that certain QFTs with Fermi surfaces might be holographically engineered by considering the family of hvLf metrics in the case $\theta=d-1$ \cite{Huijse:2011ef,Ogawa:2011bz,Dong:2012se}, as we will review in section \ref{secr3}; indeed in these cases, the HEE exhibits a logarithmic violation of the area law (note that the case $\theta=0$ precisely corresponds to AdS$_3$). Also, as observed in \cite{Dong:2012se}, the leading term in the HEE expression will not respect this \emph{law} for any value of $(d-1)\leq \theta \leq d$.\\\\
\noindent
\emph{$\circ$ Holographic entanglement entropy in higher-order gravities.\\}\noindent
In the context of holography, EE for theories dual to Einstein gravity can be computed through the Ryu-Takayanagi prescription \cite{Ryu:2006bv}\footnote{Remarkably, this prescription has been recently proven under certain conditions in \cite{Lewkowycz:2013nqa}.}. According to this, the holographic entanglement entropy (HEE) for a certain region $A$ living in the boundary of some asymptotically AdS$_{d+2}$ spacetime is given by
\be \label{RyuTaka}
S_{EG}=\underset{m\sim V}{\text{ext}}\left[\frac{\mathcal{A}(m)}{4G} \right]\, ,
\ee
where $m$ are codimension-$2$ bulk surfaces homologous to $A$ with $\partial m = \partial A$, and $\mathcal{A}(m)$ is the $d$-dimensional volume  (\emph{area}) of $m$. Hence, HEE in theories with an Einstein gravity dual is obtained by extremizing the area functional over all possible bulk surfaces homologous to $A$ whose boundary coincides with $\partial A$.

The situation changes when we start considering higher-curvature terms in the bulk Lagrangian. In such cases, the Ryu-Takayanagi prescription does not produce the correct answer for the HEE. Actually, (\ref{RyuTaka}) might be somehow regarded as a generalization of the Bekenstein-Hawking formula for the entropy of black holes \cite{Bekenstein:1973ur,Hawking:1971tu,Bardeen:1973gs}, which suggests that the expression for the EE in the presence of higher-derivative gravities might be obtained by applying the same generalization to Wald's formula, which gives the black hole entropy in this class of theories \cite{Wald:1993nt}\footnote{In (\ref{Wald}), $\mathcal{L}$ is the gravity Lagrangian, H stands for the horizon, $h_{\text{H}}$ is the induced metric on it and $\epsilon_{\mu\nu}$ is a binormal to H.}
\be \label{Wald}
S_{\text{Wald}}=\frac{1}{4G}\int_{\text{H}} d^2y \sqrt{h_{\text{H}}}\frac{\partial \mathcal{L}}{\partial R_{\mu\nu\rho\sigma}}\epsilon_{\mu\nu}\epsilon_{\rho\sigma}\, .
\ee
However, in \cite{Hung:2011xb} this guess was shown to be wrong, since this expression would produce incorrect universal terms. Alternative expressions yielding the right terms are known for Lovelock gravities \cite{Jacobson:1993xs,Hung:2011xb,Sarkar:2013swa} as well as for curvature-squared theories \cite{Fursaev:2013fta,Myers:2013lva}. Remarkably enough, a general formula for any theory involving arbitrary contractions of the Riemann tensor $\mathcal{L}(R_{\mu\nu\rho\sigma})$, which seems to satisfy several consistency checks, has been recently proposed by Dong \cite{Dong:2013qoa} (see also, e.g. \cite{Camps:2013zua,Bhattacharyya:2013jma,Bhattacharyya:2014yga,Erdmenger:2014tba})). The corresponding expressions would contain a Wald-like term as well as additional terms involving contractions of extrinsic curvatures (which vanish in the case of a Killing horizon) with second derivatives of the Lagrangian with respect to the Riemann tensor.
\\ \\
\noindent
$\circ$ \emph{Plan of the paper and motivation.\\}
In this paper we are going to study the effects of including higher-order curvature terms in the gravity Lagrangian on the HEE formula for hvLf geometries. The motivation for this study is manyfold. On the one hand, studying higher-order gravity Lagrangians in the holographic context is intrinsically interesting, given that such terms generically appear as $\alpha^{\prime}$ corrections in the appropriate String Theory embedding, corresponding to moving away from the infinitely coupled regime in the dual field theory. Secondly, as we have explained, hvLf geometries have been shown to provide interesting violations of the area law of EE for certain values of $\theta$ and, particularly interestingly, logarithmic terms for $\theta=d-1$, in whose case they have been argued to be intimately related to certain condensed matter systems. A natural question to ask is how the inclusion of higher-curvature terms will alter the structure of the HEE and whether these modifications can lead to new logarithmic terms, which might contain universal information about the dual theory (see the discussion about the UV interpretation of hvLf metrics in section \ref{conc}). Also, the expressions for HEE in higher-order Lagrangians which are known at present are restricted to a handful of theories, as explained before, and have not been \emph{proven} in general. This makes interesting to check how they perform in different situations, probing whether they produce sensible results in the different cases. An example of this is given by Gauss-Bonnet gravity in $d=2$. In such case, the HEE (which can be obtained using the so-called Jacobson-Myers (JM) functional \cite{Jacobson:1993xs})\footnote{See section \ref{secr2}.} should not change with respect to the Einstein gravity case, since the equations of motion are unchanged in this case, and any remainder of $\lambda_{GB}$ should be completely removed by including the boundary term prescribed in the JM functional.

In the next section we study the structure of divergences of HEE for a stripe in the boundary of hvLf metrics when $\theta\leq 0$, for higher-order gravities. We start with curvature-squared, for which the HEE functional is known \cite{Fursaev:2013fta}, dealing with the cases of $R^2$, Gauss-Bonnet and Ricci$^2$. We will find that a single new divergence appears in all cases, and how it cannot become logarithmic for any value of $\theta$ except for $\theta=0$, $d=1$, corresponding to the well-known AdS$_3$ case. However, extending the analysis to higher-curvature ($n$th-order) gravities we will find that new logarithmic divergences will show up for
\be \label{thetalogg}
\theta=\frac{d(d-1)}{d-2(n-1)}\, ,
\ee
provided $d<2(n-1)$. We will therefore find that an infinite family of hvLf geometries produces new logarithmic contributions to the HEE formula when these geometries are embedded in higher-curvature gravities.
For $R^2$ gravity we will be able to compute the $\mathcal{O}(\lambda_1)$ correction to the universal constant term as well. Also, in the section devoted to Gauss-Bonnet gravity, we show explicitly that the boundary term in the JM functional exactly cancels the bulk surface contribution when $d=2$, as expected.

In section \ref{secr3} we study the case $0<\theta<d$, for which we consider a UV AdS-completion of the geometry, following the steps of \cite{Ogawa:2011bz}. We will find that (\ref{thetalogg}) holds for the appearance of logarithmic contributions to the HEE, with the difference that now $d>2(n-1)$. However, both conditions together will turn out to restrict the allowed values of $\theta>0$ to the well-known case of $\theta=d-1$ \cite{Huijse:2011ef,Ogawa:2011bz,Dong:2012se}, corresponding to Einstein gravity.

In section \ref{conc} we summarize our findings, comment on possible extensions and conclude.

Finally, in appendix \ref{secr4} we consider the case in which the anisotropic scaling occurs along a spatial direction instead of time, which can be understood as a double Wick rotation of the standard hvLf geometry \cite{deBoer:2011wk,Alishahiha:2012cm}, and analyze how this changes the discussion of the previous sections. New logarithmic terms are found here for some combinations of $z$, $\theta$ and $d$.

\section{HEE for hvLf geometries in higher-curvature gravities I: $\theta\leq0$}\label{secr2}
$\circ$ \emph{Einstein gravity.\\}
Before considering higher-curvature corrections, let us start reviewing the Einstein gravity result for the HEE of hvLf geometries. We do so here for the class of metrics with $\theta\leq 0$, which we study in this section. Along this paper we will consider an entangling region $A$ consisting of a multi-dimensional infinite strip $s$ of width $l$ and infinite length $L_S\rightarrow +\infty$ (this length plays the role of an IR cut-off), $s=\left\{(t_E,r,x_1,x_2,...,x_d)\text{ s.t., }t_E=0,\right.$ $\left. \, x_1\in[-l/2,l/2],\, x_{2,...,d}\in (-L_S/2,+L_S/2)  \right\}$. As explained in the introduction, HEE for field theories dual to Einstein gravities\footnote{By this we mean theories with Lagrangians given by $\mathcal{L}=R-2\Lambda+\mathcal{L}_{\text{other fields}}$.} can be computed using the Ryu-Takayanagi prescription \cite{Ryu:2006bv}
\be
S_{EG}=\frac{1}{4G}\int_m d^dx\sqrt{g_m}\, ,
\ee
where $m$ is the bulk surface homologous to $A$, with $\partial m=\partial A$, which extremizes the above functional, and $g_m$ is the determinant of the induced metric on $m$.

The translational symmetry of the strip along the directions $2,...,d$ allows us to parametrize the entangling surface $m$ as $r=h(x_1)$. For our hvLf geometry (\ref{hvLf}), the induced metric on such a surface reads
\begin{equation}\label{indS}
ds^2_{m}=L^2 h^{\frac{2(\theta-d)}{d}}\left[\left[1+\dot{h}^2 \right]dx_1^2+ d\vec{x}^2_{(d-1)}\right]\, ,
\end{equation}
where $ d\vec{x}^2_{(d-1)}\equiv dx_2^2+...+dx_{d}^2$. Using this expression and the fact that $m$ must be mirror symmetric with respect to the plane $x_1=0$, we find
\be \label{eeeg}
S_{EG}=\frac{L ^d L_S^{(d-1)}}{2G}\int^{l/2}_0 dx_1\,h^{(\theta-d)} \sqrt{1+\dot{h}^2}\, ,
\ee
The \emph{Lagrangian} does not depend explicitly on $x_1$, so we have a conserved quantity
\be\label{he}
h_*^{(\theta-d)}=\frac{h^{(\theta-d)}}{\sqrt{1+\dot{h}^2}}\, ,
\ee
where $h_*$ is the turning point of the surface, in which $\dot{h}|_{h_*}=0$. Substituting this expression in (\ref{eeeg}), we find
\be
S_{EG}=\frac{L ^d L_S^{(d-1)} h_*^{(\theta-d+1)}}{2G}\int^{1}_{\delta/h_*} \frac{u^{(\theta-d)}du}{\sqrt{1-u^{2(d-\theta)}}}\, ,
\ee
where we made the change of variable $u=h/h_*$ and introduced the UV cut-off $(h(x_1)\rightarrow \delta) \leftrightarrow (x_1\rightarrow \pm l/2)$. The turning point is related to the strip width through
\begin{equation}\label{tpEG}
\frac{l}{2}=\int_0^{l/2}dx_1=h_* \int_0^{1}\frac{u^{(d-\theta)}\, du}{\sqrt{1-u^{2(d-\theta)}}}=h_*\frac{\sqrt{\pi}\Gamma\left(\frac{1+d-\theta}{2(d-\theta)}\right)}{\Gamma\left(\frac{1}{2(d-\theta)}\right)}  \, .
\end{equation}
These two integrals allow us to obtain the final expression for the entanglement entropy of the strip
\be \label{eeegb}
S_{EG}=\frac{L^d L_S^{(d-1)} }{2G (d-\theta-1)} \left[\delta^{-(d-\theta-1)}-(l/2)^{(\theta-d+1)}   \left[\frac{\sqrt{\pi} \Gamma\left(\frac{1+d-\theta}{2(d-\theta)}\right)}{\Gamma\left(\frac{1}{2(d-\theta)}\right)} \right]^{(d-\theta)}\right]\, .
\ee
This is the beautiful formula found in \cite{Dong:2012se}. As we can see, the scaling behavior of the HEE gets modified with respect to the AdS$_{d+2}$ case \cite{Ryu:2006ef} by factors with dimensions of (length)$^{\theta}$. In particular, we find a corrected exponent for the divergent term of order
\be
\mathfrak{B}_0\equiv d-\theta-1\, .
\ee
Of course, $\mathfrak{B}_0$ is always positive for $\theta< 0$. One can introduce an intermediate scale $r_F$ as explained in the introduction, which would modify the factors $\delta^{\theta}\rightarrow (\delta/r_F)^{\theta}$ and $(l/2)^{\theta}\rightarrow (l/(2r_F))^{\theta}$. When $\theta=0$, we recover the usual AdS$_{d+2}$ expression \cite{Ryu:2006ef}
\be \label{eeegb2}
S_{EG}=\frac{L^d L_S^{(d-1)} }{2G (d-1)} \left[\delta^{-(d-1)}-(l/2)^{(1-d)}   \left[\frac{\sqrt{\pi} \Gamma\left(\frac{1+d}{2d}\right)}{\Gamma\left(\frac{1}{2d}\right)} \right]^{d}\right]\, ,
\ee
which in the limit case of $d=1$, corresponding to AdS$_3$, yields a logarithmic divergence
\be \label{eeegb2}
S_{EG}=\frac{L  }{2G} \log\left[ \frac{l}{\delta} \right]\, .
\ee
It is well-known that hvLf geometries can produce logarithmic terms in the HEE for $\theta=d-1$. However, given that these cases correspond to metrics with $0<\theta<d$ for $d\geq 2$, we will review them in section \ref{secr3}, along with the corresponding new higher-order terms.
\\ \\
\noindent
$\circ$ \emph{Higher-curvature corrections to HEE.}\\
We are interested now in considering higher-order curvature corrections to the bulk action and see how they affect the HEE expression for hvLf geometries. In general, the gravitational action will be given by Einstein's gravity plus an (infinite) sum of higher-curvature terms with small coupling constants (otherwise, the semiclassical approximation would not make sense)
\begin{equation}\label{actr2}
\mathcal{I}_g=\frac{1}{16\pi G}\int d^{d+2}x\sqrt{g}\left[R+\frac{d(d+1)}{\tilde{L}^2}+\tilde{L}^2\left[\lambda_1  R^2+\lambda_2  R_{\mu\nu}R^{\mu\nu}+\lambda_3  R_{\mu\nu\rho\sigma}R^{\mu\nu\rho\sigma}\right]+\tilde{L}^4\mathcal{O}(R^3) \right] \, ,
\end{equation}
being $\tilde{L}$ a length scale which would coincide with the AdS$_{d+2}$ radius $L$ for Einstein gravity, but would be different in general otherwise, and $\lambda_{1,2,3,...}$ dimensionless couplings.

The next step would correspond now to choose some \emph{matter} content and solve the equations of motion for the corresponding fields trying to determine if our hvLf family of metrics (\ref{hvLf}) can be embedded into the theory. The case of curvature-squared gravity was studied in \cite{O'Keeffe:2013nha}, where the authors consider an EMD system with general curvature-squared corrections. For our purposes, it suffices to recall the fact that hvLf geometries are indeed solutions of the corresponding equations of motion, and are expected to appear as well as solutions to similar EMD gravities with even higher-curvature corrections. Another interesting piece of information we can extract from \cite{O'Keeffe:2013nha} is the fact that the NEC arising in a general EMD curvature-squared gravity reduces in general to a pair of conditions on $(z, \theta)$ and the couplings of the new terms, plus the well-known NEC of the Einstein gravity case \cite{Dong:2012se}
\begin{eqnarray}
(z-1)(z-\theta+d)\geq 0\, ,\\
(d-\theta)(d(z-1)-\theta)\geq 0\, ,
\end{eqnarray}
which in the case under consideration in this paper, i.e., $d>\theta$, reduces to the condition $z\geq 1$. From now on, we restrict ourselves to this case, although as we will see, our results would not get modified for $z<1$ since $z$ will not appear in the exponents of the different terms in the HEE expressions for our hvLf geometries\footnote{The situation will change in appendix \ref{secr4}, where we will consider a doubly Wick-rotated version of (\ref{hvLf}).}.

Unfortunately, computing HEE in general higher-curvature gravities is a very hard task at present because Dong's recipe \cite{Dong:2013qoa} turns out to be difficult to apply in most cases, with some exceptions: Lovelock \cite{Jacobson:1993xs,Hung:2011xb}, curvature-squared \cite{Fursaev:2013fta} and $f(R)$ gravities \cite{Wald:1993nt,Dong:2013qoa}. Nevertheless, making use of the results found in curvature-squared gravity plus some general arguments, which we will discuss in a moment, we will to try to say something about the structure of divergences of the HEE in any higher-curvature gravity for our hvLf geometries.

There are two steps one needs to take in order to successfully obtain the HEE expression in any higher-curvature gravity for any background, assuming the HEE functional is known. The first is extremizing such a functional, whereas the second corresponds to evaluating the on-shell integral. The first one is undeniably harder in general, since the equations of motion we pretend to solve will usually be of high order in derivatives, and very non-linear. However, we can note the following: in the HEE expression we will find in general a sum of divergent terms coming from the on-shell evaluation of the integral near the boundary, plus a constant term related to the bulk contribution. In geometries in which the higher the order of the curvature term the faster it goes to zero in the UV, we will find an expression consisting of a leading Einstein gravity divergence plus possible subleading divergences coming from the higher-order terms, plus a constant term. The question is now how the fact that the entangling surface is different in higher-order gravities with respect to the Einstein gravity case affects the HEE expression, given that the functional we need to extremize is different. We expect the surface to be significantly different away from the UV, where the new terms become large, producing therefore new corrected constant terms. However, as we approach the boundary, where the divergences are to appear, the higher-order terms will die out, and the shape of the entangling surface should not differ much from the Einstein gravity one. This is analogous to computing the area for different surfaces sharing boundary with the extremal area one, $m$. The result will of course differ, but the order of the divergences will be the same as the one found for $m$. Thus, it is reasonable to expect that the new divergent terms (if any) appearing in the HEE expression for higher-curvature terms will be produced from the evaluation of the on-shell integral using the surface which extremizes the area functional of Einstein gravity, without having to find the surface which extremizes the new functional. In other words, the new entangling surface should not change the structure of divergences with respect to the one with extremal area and this has two interesting consequences. First, we can identify the order of the divergences of higher-order gravity terms using the extremal area surface, and second, every new divergence will appear at order $\mathcal{O}(\lambda)$ in the corresponding gravitational coupling.
Therefore, any term of order $\mathcal{O}(\lambda^2)$ or higher will appear next to a constant, arising from the bulk contribution to the integral.

At this point it is convenient to stress that the study of the structure of divergences of the HEE is physically motivated by the fact that it allows us to determine the dependence of the different terms with the size of the entangling region. In particular, we can use this to check if the area law holds, unveil the presence of universal terms, etc.

Let us now turn to the real calculations. We are going to study in full detail the case of $R^2$ gravity, in which we will be able to compute the corrected extremal surface. This will allow us to illustrate how the above argument works, and use it to compute the structure of divergences for general curvature-squared gravities, including the more involved cases of Gauss-Bonnet and Ricci$^2$ gravities. We will finish this section showing how the results found for these theories allow us to conjecture the form of all divergences in any higher-order curvature gravity for our hvLf metrics. Let us start with curvature-squared gravities.

%

\subsection{$R^2$ gravity}
The most general curvature-squared gravity action can be written in terms of three contractions involving the Riemann tensor. These can be chosen to be
\begin{equation}\label{acr2}
\mathcal{I}_{\text{curv}^2}=\frac{1}{16\pi G}\int d^{d+2}x\sqrt{g}\left[R+\frac{d(d+1)}{\tilde{L}^2}+\tilde{L}^2 \left[\lambda_1 R^2+\lambda_2 R_{\mu\nu}R^{\mu\nu}+\lambda_{GB}\mathcal{X}_4\right] \right] \, ,
\end{equation}
where $\mathcal{X}_4=R^2-4 R_{\mu\nu}R^{\mu\nu}+R_{\mu\nu\rho\sigma}R^{\mu\nu\rho\sigma}$ is the Gauss-Bonnet term, which in four bulk dimensions corresponds to the Euler density of the spacetime manifold.

\begin{figure}[h]\centering
   \includegraphics[scale=0.5]{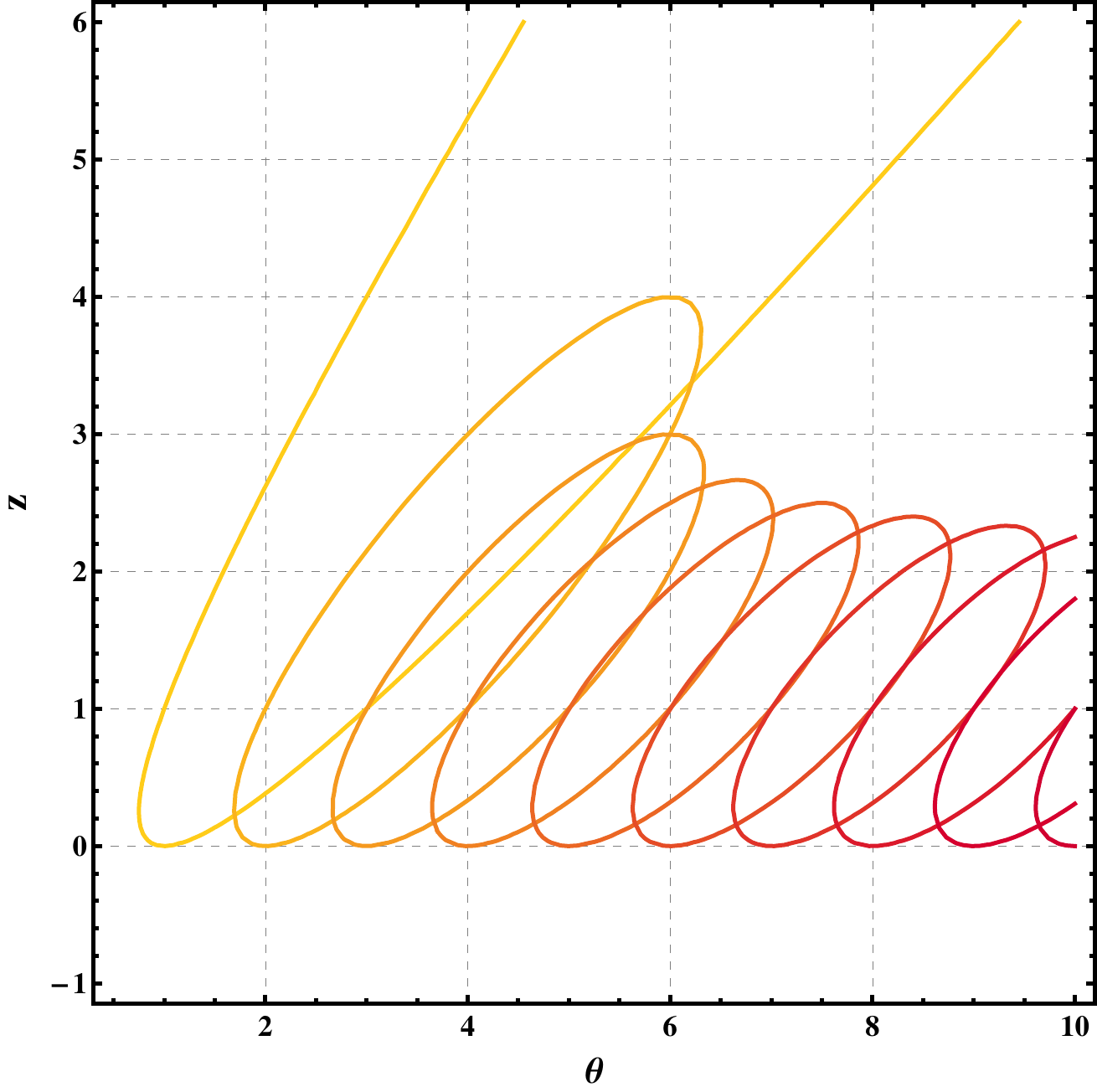}
\caption{Curves $(\theta,z)$ for which the Ricci scalar of hvLf metrics vanishes. $d=1$ is depicted in yellow, whereas darker lines correspond to $d=2,3,...$ }
\label{riccizero}

\end{figure}

In the case of $R^2$ gravity, the HEE functional\footnote{The functional proposed by \cite{Fursaev:2013fta} for the HEE of curvature-squared gravities has been used in several works, including \cite{Bhattacharyya:2013gra,Alishahiha:2013dca,Alishahiha:2013zta}.} is given by \cite{Fursaev:2013fta}
\begin{eqnarray}
\label{seer2}\hspace{-1cm}
S_{R^2}=\frac{1}{4G} \int_{m} d^dx \sqrt{g_{m}}\left[1+2\lambda_1 \tilde{L}^2 R\right]\, .
\end{eqnarray}
For our hvLf metrics (\ref{hvLf}) the Ricci scalar reads
\be
R=\kappa \frac{r^{-2\theta/d}}{\tilde{L^2}}\, ,
\ee
where we have defined the constant
\be
\kappa \equiv -\frac{2 \tilde{L}^2}{L^2}\left[z^2+ z d+\frac{d+1}{2}\left[d-2\theta-\frac{\theta}{d}(2z-\theta) \right]\right]\, .
\ee
As a curiosity, there are certain combinations of $(z,\theta)$ for which $\kappa$ vanishes, meaning that the $R^2$ contribution identically vanishes, and does not produce any correction at all with respect to the Einstein gravity result. The corresponding curves for which this happens are shown in Figure \ref{riccizero}.
Leaving this case aside, the expression for the entanglement entropy of the strip becomes, using (\ref{indS})
\be \label{eer2}
S_{R^2}=\frac{L ^d L_S^{(d-1)}}{2G}\int^{l/2}_0 dx_1\,h^{(\theta-d)} \sqrt{1+\dot{h}^2}\left[1+2 \kappa \lambda_1 h^{-2\theta/d}\right]\, .
\ee
Since the functional does not depend on $x_1$ explicitly, there is again a first integral which we can use to write the expression for $\dot{h}$ in terms of $h$. We have
\be \label{egsurf}
\sqrt{1+\dot{h}^2}=\frac{f(h) h^{(\theta-d)}}{f(h_*) h_*^{(\theta-d)}} \, , \,\,\, \text{with} \,\,\, f(x)\equiv \left[1+2 \kappa \lambda_1 x^{-2\theta/d} \right] \, ,
\ee
where $h_*$ is again the turning point of the surface, characterized by $\dot{h}|_{h_*}=0$. We can use this relation to rewrite (\ref{eer2}) in terms of $u \equiv h/h_*$ as
\be \label{eer3}
S_{R^2}=\frac{L ^d L_S^{(d-1)} h_*^{\theta-d+1}}{2G}\int^{1}_{\delta/h_*} du\,\frac{u^{(\theta-d)} f(u h_*)}{\sqrt{1-u^{2(d-\theta)}\frac{f(h_*)^2}{f(u h_*)^2}}}\, ,
\ee
where we have introduced again an ultraviolet cut-off $h\rightarrow \delta$ to account for the divergent terms. Note that despite the intricated appearance of the integrand it is already possible at this level to keep track of those divergences. Indeed we can study its behaviour in the limit $u \rightarrow 0$
\be \label{limit}
\lim_{u \to 0} \frac{u^{(\theta-d)} f(u h_*)}{\sqrt{1-u^{2(d-\theta)}\frac{f(h_*)^2}{f(u h_*)^2}}}=u^{(\theta-d)} \left[1+2\kappa \lambda_1 (u h_*)^{-2\theta/d} \right] \left[1+\mathcal{O}\left( u^{2(d-\theta)} \right) \right] \, ,
\ee
so the terms with a negative power in $u$, and therefore those resulting into divergences, arise from the product $u^{(\theta-d)}\left[1+2\kappa \lambda_1 (u h_*)^{-2\theta/d} \right]$. This agrees with what we anticipated in our previous discussion: had we taken the Einstein gravity surface (\ref{indS}), and computed the HEE integral (\ref{eer2}), we would have found the same divergent terms. It is also important to stress that this expression is valid for any value of the coupling $\lambda_1$, so if we expanded in powers of $\lambda_1$, the only divergence would appear at order $\mathcal{O}(\lambda_1)$, as anticipated.  Taking into account (\ref{limit}) we find that the entanglement entropy is of the form
\be \label{eer4}
S_{R^2}=\frac{L ^d L_S^{(d-1)}}{2G} \left[ \frac{1}{\mathfrak{B}_0} \delta^{-\mathfrak{B}_0}+\frac{2 \kappa \lambda_1 }{\mathfrak{B}_1} \delta^{-\mathfrak{B}_1} \right] + S_0 \, ,
\ee
with
\begin{eqnarray}
\mathfrak{B}_0&\equiv& d-\theta-1\, , \\
\mathfrak{B}_1&\equiv&\mathfrak{B}_0+\frac{2\theta}{d}\, ,
\end{eqnarray}
and $S_0$ being a constant term which we will discuss later. As we can see, the inclusion of the $R^2$ term introduces a new divergence in the HEE. This contribution is not dominant, and the leading divergence is again the Einstein gravity, one as expected. It is also impossible to produce a logarithmic divergence from this term, since this would correspond to $\theta=\frac{d(d-1)}{(d-2)}$, which is larger than $0$ for any $d> 1$. An exception is $d=1$, $\theta=0$, which would correspond to AdS$_3$, for which both $\mathfrak{B}_0$ and $\mathfrak{B}_1$  would be logarithmic. In the special case of Lifshitz geometries, $\theta=0$, the Ricci scalar is constant and the entanglement entropy diverges as
\be \label{eer5}
S_{R^2}|_{\theta=0}=\left(1+2 \kappa|_{\theta=0} \lambda_1 \right) S_{EG}|_{\theta=0} \, ,
\ee
where $S_{EG}|_{\theta=0}$ is just the HEE for a strip in AdS$_{d+2}$ (recall that, although $z\neq 1$ in general, the dynamical exponent does not enter into the HEE expression for Einstein gravity), which can be read from (\ref{eeegb}), and
\be
\kappa|_{\theta=0}=-\frac{2 \tilde{L}^2}{L^2}\left[z^2+ z d+\frac{d(d+1)}{2}\right]\, .
\ee
As we can see, the dynamical exponent does appear in the HEE formula (through $\kappa$) when we consider this curvature-squared contribution, as opposed to the Einstein gravity case\footnote{The fact that a Lifshitz geometry ($\theta=0$) produced an unaltered HEE with respect to the AdS case for Einstein gravity was first observed in \cite{Solodukhin:2009sk}.}. However, it does not contribute to the exponents of the divergences, and it will not do so for any higher-curvature gravity, simply because the induced metric on any entangling surface extremizing the corresponding functional will not depend on $z$ in general, given that it only appears in the $g_{tt}$ component of the hvLf metric (\ref{hvLf}). In order to make $z$ appear in the exponents of the HEE terms, we need to consider an anisotropic scaling of a spatial coordinate instead of time. This will be studied in appendix \ref{secr4}. The appearance of the new divergence $\delta^{-\mathfrak{B}_1}$ is a distinctive feature of hvLf geometries: for AdS or even Lifshitz geometries, the inclusion of additional higher-curvature terms in the bulk action just shifts the coefficient in front of $\delta^{-\mathfrak{B}_0}$, without producing any new divergent term.

Coming back to $R^2$ gravity, in order to extract information about the finite term $S_0$ in (\ref{eer4}) we are going to consider the case $\lambda_1<<1$ (which is a reasonable assumption as we are considering the higher-curvature terms to be corrections to the leading Einstein gravity action), so we can Taylor-expand around $\lambda_1=0$. We do so in the expression for the entanglement entropy up to order $\lambda_1$ and perform the integration afterwards. The result reads
\be\label{eer2b}
S_{0}=-\frac{L ^d L_S^{(d-1)} }{2G} \left\{ \frac{G_0 h_*^{-\mathfrak{B}_0}}{\mathfrak{B}_0}+2 \kappa \lambda_1 h_*^{-\mathfrak{B}_1}\left[\frac{G_0}{(\mathfrak{B}_0+1)} +G_1\left[\frac{1}{\mathfrak{B}_1}-\frac{1}{(\mathfrak{B}_0+1)}\right]\right]\right\}+\mathcal{O}(\lambda_1^2)\, ,
\ee
where we defined the constants
\be \label{gs}
G_0\equiv\frac{\sqrt{\pi}\Gamma\left(\frac{\mathfrak{B}_0+2}{2(\mathfrak{B}_0+1)}\right)}{\Gamma\left(\frac{1}{2(\mathfrak{B}_0+1)}\right)} \, , \,\, \,G_1\equiv\frac{ \sqrt{\pi} \Gamma\left(\frac{2
+2\mathfrak{B}_0-\mathfrak{B}_1}{2(\mathfrak{B}_0+1)}\right)     }{ \Gamma\left(\frac{1+\mathfrak{B}_0-\mathfrak{B}_1}{2(\mathfrak{B}_0+1)}\right)}\, .
\ee
The turning point $h_*$ is in this case related to the strip width through
\be
\frac{l}{2}=\int_0^{l/2}dx_1=h_* \int_0^{1}\frac{f(h_*) u^{(d-\theta)}\, du}{f(u h_*)\sqrt{1-u^{2(d-\theta)}\frac{f(h_*)^2}{f(u h_*)^2}}} \, .
\ee
At first order in $\lambda_1$, we can perform the integral and invert the expression to find
\be
h_*=\frac{l/2}{G_0}\left[1+\frac{2\kappa\lambda_1}{(\mathfrak{B}_0+1)}\left[\frac{l/2}{G_0}\right]^{(\mathfrak{B}_0-\mathfrak{B}_1)}\left[1-\frac{G_1}{G_0} \right] \right]\, .
\ee
Substitution into (\ref{eer2b}) leads to a kind simplification, and the full entanglement entropy expression at this order is finally given by
\be \label{sr2}
S_{R^2}=\frac{L ^d L_S^{(d-1)} }{2G} \left\{\frac{\delta^{-\mathfrak{B}_0}}{\mathfrak{B}_0}-\frac{(l/2)^{-\mathfrak{B}_0 }G_0^{\mathfrak{B}_0}G_0}{\mathfrak{B}_0}+2\kappa \lambda_1\left[\frac{\delta^{-\mathfrak{B}_1}}{\mathfrak{B}_1}-\frac{(l/2)^{-\mathfrak{B}_1 }G_0^{\mathfrak{B}_1}G_1}{\mathfrak{B}_1} \right] \right\}+\mathcal{O}(\lambda_1^2) \, .
\ee
This expression is exact at linear order in $\lambda_1$. The Einstein gravity result, given by the first two terms, is corrected by a divergent plus a constant term at first order, plus a constant contribution of order $\mathcal{O}(\lambda_1^2)$.

\subsection{Gauss-Bonnet gravity}
Let us now turn to the case of Gauss-Bonnet gravity. The HEE functional for this theory was proposed in \cite{Hung:2011xb} and, as we mentioned, corresponds to a particular case of the JM functional, suitable for Lovelock gravities. Including the boundary term, which we will make use of for $d=2$, the expression reads
\begin{equation}
\label{seeGB88}
S_{GB}=\frac{1}{4G} \int_{m} d^{d}x \sqrt{g_m}\left[1+2\lambda_{GB}\tilde{L}^2 \mathcal{R}_m \right]+\frac{\lambda_{GB}\tilde{L}^2}{G}\int_{\partial m} d^{d-1}y \sqrt{g_{\partial m}} \mathcal{K}\, ,
\end{equation}
where $\mathcal{R}_m$ is the Ricci scalar of $m$, $\partial m$ is the $(d-1)$-dimensional boundary of $m$, $h_{\partial m}$ stands for the determinant of the induced metric on $\partial m$, and $\mathcal{K}$ is the trace of its extrinsic curvature.

In the case of our hvLf geometries, the Ricci scalar of the induced metric on $m$ (\ref{indS}) reads
\be \label{Rm}
\mathcal{R}_m=\frac{(d-1)(d-\theta) h^{-2\theta/d}}{(1+\dot{h}^2)^2L^2}\left[\left(\dot{h}^2+\dot{h}^4\right)\left(\frac{(d-2)\theta}{d^2}-1 \right) +\frac{2 h \ddot{h}}{d}\right]\, .
\ee
As we can see, it identically vanishes for $d=1$, which was expectable since the Gauss-Bonnet term $\mathcal{X}_4$ is identically zero in 3D gravity\footnote{The same would occur for $d=\theta$, so no corrections to HEE are produced by this term in such a limit case.}.

The way to proceed now is again trying to extremize (\ref{seeGB88}) and evaluate the \emph{on-shell} integral. The simplest case and, at the same time, one of singular interest, is given by $d=2$. There, the Gauss-Bonnet contribution reduces to a boundary term, and does not modify the gravitational equations of motion. From the HEE perspective, the integral of the Ricci scalar of a 2D surface embedded in a certain manifold (which is precisely the expression we have here) is proportional to its Euler characteristic, which is a topological quantity, independent of the geometry of $m$. Therefore, when $d=2$ we expect the entangling surface to be the same as in Einstein gravity and the Gauss-Bonnet bulk contribution $\propto \int \mathcal{R}_m$ to be cancelled by the boundary term involving the integral of the extrinsic curvature of $\partial m$. Let us explicitly show that this is indeed the case for hvLf geometries.

It is straightforward to check that the equations of motion for $h(x_1)$ do not get modified, and we have the very same first integral as in the Einstein gravity case (\ref{he}), which we rewrite here for convenience
\be \label{hed2}
h_*^{(\theta-2)}=\frac{h^{(\theta-2)}}{\sqrt{1+\dot{h}^2}}\, .
\ee
The Ricci scalar on $m$ simplifies to
\be
\mathcal{R}_m=\frac{(\theta-2)}{h_*^{\theta}L^2}\left[u^{-\theta}-(\theta-1)u^{(4-3\theta)} \right]\, ,
\ee
where we have used again $u\equiv h/h_*$. We can now compute the integral involving the bulk terms in (\ref{seeGB88}). The result is a sum of the Einstein gravity term (\ref{eeegb}) and the following divergence
\begin{equation}
\label{seeGB5}
\frac{1}{4G} \int_{m} d^{d}y \sqrt{g_m}\left[2\lambda_{GB}\tilde{L}^2 \mathcal{R}_m \right]=\frac{(2-\theta)\tilde{L}^2 L_S\lambda_{GB}}{2G}\frac{1}{\delta}\, .
\end{equation}
Interestingly, the exponent of the divergence does not depend on $\theta$. In order to verify the cancellation of this term with the boundary one, we need to compute the metric induced on $\partial m$, and the trace of the extrinsic curvature of such boundary understood as an embedding on $m$. $\partial m $ is characterized by $h\rightarrow \delta$, $x_1=\text{const.}$ We find, after some algebra
\begin{eqnarray} \displaystyle
\sqrt{g_{\partial m}}&=&L \delta^{(\frac{\theta-2}{2})}\, ,\\ \notag
\mathcal{K}_{\partial m}&=&\frac{(\theta-2)}{2}\frac{\delta^{-\frac{\theta}{2}}}{L}\, ,
\end{eqnarray}
and hence
\be
\frac{\lambda_{GB}\tilde{L}^2}{G}\int_{0}^{L_S} dx_2 \sqrt{g_{\partial m}} \mathcal{K}=\frac{(\theta-2)\tilde{L}^2 L_S\lambda_{GB}}{2G}\frac{1}{\delta}\, .
\ee
As we can see, this contribution exactly cancels the intrinsic curvature contribution of (\ref{seeGB5}), as expected.

In the case $d>2$ things get much more involved. The functional we pretend to extremize contains derivatives of $h(x_1)$ up to order two, so no first integral is available now. Similarly, although the equations of motion are second-order as well, and not fourth-order as one would expect for a random second-order gravity\footnote{Recall Gauss-Bonnet is a particular Lovelock gravity, which is the most general family of higher-order gravity theories in any dimension with second-order equations of motion.}, they turn out to be impossible to treat analytically. However, as we argued before we do not need to obtain the surface extremizing (\ref{seeGB88}) in order to obtain the divergent terms in the HEE expression (although we would if we wanted to provide the corresponding corrected constant terms). Indeed, let us use (\ref{he}) to compute the divergences produced by the bulk integral in (\ref{seeGB88}). Following the same steps as for $R^2$ gravity we find\footnote{For the case $d=3$, the appearance of $\mathfrak{B}_1$ in Gauss-Bonnet was anticipated in \cite{Kulaxizi:2012gy}.}
\be \label{sr3}
S_{GB}=\frac{L ^d L_S^{(d-1)} }{2G} \left\{\frac{\delta^{-\mathfrak{B}_0}}{\mathfrak{B}_0}-\frac{(l/2)^{-\mathfrak{B}_0 }G_0^{\mathfrak{B}_0}G_0}{\mathfrak{B}_0}+\xi \lambda_{GB}\left[\frac{\delta^{-\mathfrak{B}_1}}{\mathfrak{B}_1}+ c_{1,GB} \right] \right\}+\mathcal{O}(\lambda_{GB}^2) \, ,
\ee
where now
\be
\xi\equiv \frac{\tilde{L}^2}{L^2}(d-1)(d-\theta)\, ,
\ee
and $c_{1,GB}$ is a constant term that should be computed using the entangling surface extremizing (\ref{seeGB88}). As we can see, the expression is completely analogous to the one found for $R^2$ gravity (\ref{sr2}): added to the Einstein gravity contribution we find a single divergence of the same order as the one encountered in that case plus a constant correcting the universal term. The fact that the divergences produced by $R^2$ and Gauss-Bonnet gravities match is not trivial, given that in the first case we are simply adding a term scaling as $\sim u^{-2\theta/d}$ (see (\ref{eer2})) to the ``1'' of Einstein gravity in the HEE integral, whereas for Gauss-Bonnet we find two terms when we substitute $\dot{h}(h)$ and $\ddot{h}(h)$ in (\ref{Rm}) and (\ref{seeGB88})): one scaling like the $R^2$ one, plus another one going as $\sim u^{-2\theta/d+2(d-\theta)}$ which, however, does not produce divergences when $\theta\leq 0$. In this case, the dynamical exponent does not appear in the curvature-squared contribution, simply because it does not appear in the pull-back metric on $m$ and, as a consequence, in $\mathcal{R}_m$. Let us see what happens for our last curvature-squared theory: Ricci-squared gravity.

\subsection{$R_{\mu\nu}R^{\mu\nu}$ gravity}
For this theory, the entanglement entropy functional reads \cite{Fursaev:2013fta}
\be\label{ricci2}
S_{\text{Ricci}^2}=\frac{1}{4G} \int_m d^d x \sqrt{g_m} \left[ 1+ \lambda_2 \tilde{L}^2 \left( R_{(\hat{a})}\,^{(\hat{a})}-\frac{1}{2} K^{(\hat{a})} \, ^2 \right) \right] \, .
\ee
In this expression, the first term stands for the contraction of the Ricci tensor associated to the spacetime metric with the two mutually orthogonal unit vectors normal to the entangling surface $m$, $n_{(\hat{a})}$, $\hat{a}=1,2$ according to
\be
R_{(\hat{a})}\,^{(\hat{a})}\equiv R_{\mu\nu} n^{\mu}_{(\hat{a})} n^{\nu}_{(\hat{b})} \delta^{(\hat{a})(\hat{b})}\, .
\ee
The second term is the sum of the squares of the two extrinsic curvatures of $m$
\be
K^{(\hat{a})}_{\mu\nu}=\nabla_{\mu} n^{(\hat{a})}_{\nu}\, ,
\ee
associated to those two vectors
\be
K^{(\hat{a})} \, ^2\equiv g^{\mu\nu}g^{\rho\sigma} K^{(\hat{a})}_{\mu\nu} K^{(\hat{b})}_{\rho\sigma} \delta_{(\hat{a})(\hat{b})}\, .
\ee
For the hvLf metrics (\ref{hvLf}), the two vectors normal to the entangling surface $m$ associated to our strip are given by
\be n_{(1)}=\frac{r^{z-\theta/d}}{L}\partial_t\, , \, \,\, n_{(2)}=\frac{r^{1-\theta /d}}{L \sqrt{1+\dot{h}^2}} \left( \partial_r -\dot{h} \partial_{x_1} \right)\, .
\ee
Making use of this we can evaluate the above expressions to get
\begin{eqnarray}\hspace{-0.3cm}
R_{(\hat{a})}\,^{(\hat{a})}-\frac{1}{2} K^{(\hat{a})} \, ^2&=&  \frac{h^{-2\theta /d}}{d^2 L^2} \left[  d(d+d z-2 \theta ) (\theta-d-z )+\frac{ d \left[\theta ^2+d ((1-z) z-\theta )\right] }{1+\dot{h}^2}\right.\\ \notag &-&\left.\frac{\left[( \theta(d+1)-d(d+z) ) (1+\dot{h}^2)+d h \ddot{h}\right]^2}{2\left[1+\dot{h}^2\right]^3}\right] \, .
\end{eqnarray}
Following our previous steps, we can make use of (\ref{he}) to determine the divergences in the HEE for this theory. The result is
\be \label{sr4}
S_{\text{Ricci}^2}=\frac{L ^d L_S^{(d-1)} }{2G} \left\{\frac{\delta^{-\mathfrak{B}_0}}{\mathfrak{B}_0}-\frac{(l/2)^{-\mathfrak{B}_0 }G_0^{\mathfrak{B}_0}G_0}{\mathfrak{B}_0}+\gamma \lambda_{2}\left[\frac{\delta^{-\mathfrak{B}_1}}{\mathfrak{B}_1}+ c_{1,\text{Ricci}^2} \right] \right\}+\mathcal{O}(\lambda_2^2) \, ,
\ee
where now
\be
\gamma \equiv \frac{\tilde{L}^2}{L^2}\frac{(d+d z-2 \theta ) (\theta-d-z )}{d}\, ,
\ee
and $c_{1,\text{Ricci}^2}$ is the correction to the constant term at first order in $\lambda_{2}$. Again, we find the same kind of term as in the two previous cases. In light of this, we conclude that $\mathfrak{B}_1=2\theta/d+d-\theta-1$ is the only new divergent term produced at the level of curvature-squared gravities when $\theta<0$. As we already said, this means that no additional logarithmic divergences can appear at this order of curvature for this class of metrics.

\subsection{Higher-curvature gravities and new logarithmic terms}
In the previous subsections we have studied the structure of terms of HEE for general curvature-squared gravities in the case of an entangling region $A$ consisting of a strip in the boundary of hvLf metrics with $\theta \leq 0$. The result is that, in spite of the different terms appearing for the distinct HEE functionals in the various curvature-squared theories, we find that one single additional divergent term appears. This might suggest that if we moved on and considered even higher curvature gravities, one single additional divergence would appear at each order in curvature (this would mean, e.g., that the 10 independent curvature-cubed gravities \cite{Decanini:2007zz}, with their different corresponding functionals would give rise to the same single divergent term, and so on). Although this conjecture seems to ask for stronger evidence, it is important to notice that at the curvature-squared gravities level we are already considering the two kinds of terms that are expected to appear in the HEE functional at all orders in curvature \cite{Dong:2013qoa}, namely: contractions of curvature bulk tensors with normal vectors to the entangling surface $m$, and contractions of extrinsic curvatures of $m$ with bulk tensors. If our conjecture was right, we could extract the divergent term common to all theories at each order in curvature by computing the HEE expression for the simplest higher-order gravity in each order. This is, of course, $R^n$ gravity.

\begin{figure}[h]\centering
   \includegraphics[scale=1]{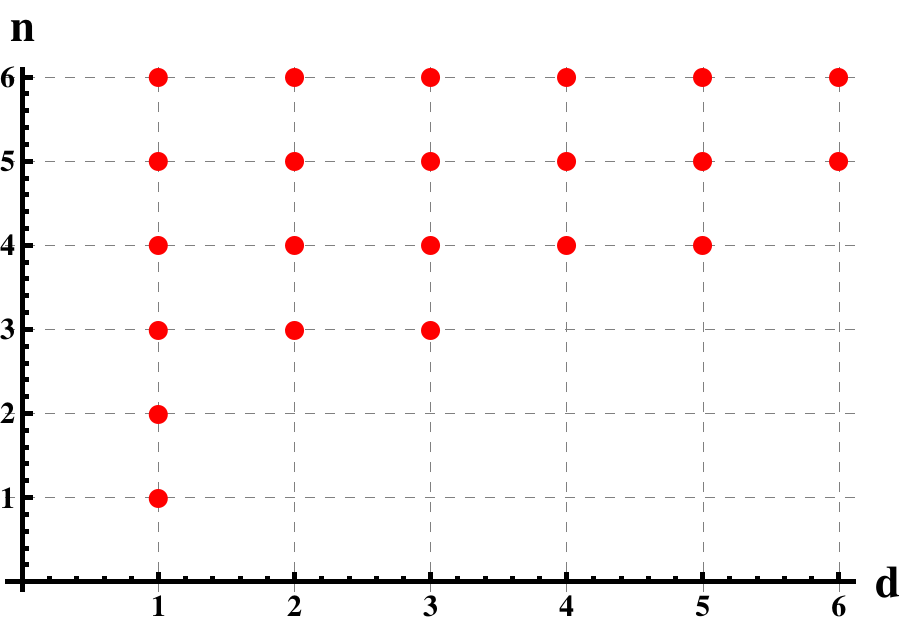}
\caption{Values of $n$ and $d$ for which the corresponding $R^n$ gravities produce terms including a logarithmic dependence on $l$ for certain values of $\theta \leq 0$. The graph extends to the $n>6$, $d>6$ region in an obvious way.}
\label{logs1}

\end{figure}

For an $R^n$ gravity or, more in general, for an $f(R)$ gravity
\begin{equation}\label{acfr}
I_{f(R)}=\frac{1}{16\pi G}\int d^{d+2}x\sqrt{g}\left[R+\frac{d(d+1)}{\tilde{L}^2}+\lambda_{f(R)} f(R) \right] \, ,
\end{equation}
(where $\lambda_{f(R)}$ is now a dimensionful coupling), the HEE functional is known to be \cite{Dong:2013qoa}
\begin{eqnarray}
\label{seefr}
S_{f(R)}=\frac{1}{4G} \int_{m} d^2x \sqrt{g_{m}}\left[1+\lambda_{f(R)} \frac{df(R)}{dR} \right]\, ,
\end{eqnarray}
and so for $f(R)=R^n$, $\lambda_{f(R)}=\lambda_{R^n}\tilde{L}^{2(n-1)}$ and
\begin{eqnarray}
\label{seern}
S_{R^n}=\frac{1}{4G} \int_{m} d^2x \sqrt{g_{m}}\left[1+n \lambda_{R^n}\tilde{L}^{2(n-1)} R^{(n-1)} \right]\, .
\end{eqnarray}
We can actually extremize this functional and find the HEE expressions following exactly the same steps as in the case of $R^2$. The result is
\be \label{srn}
S_{R^n}=\frac{L ^d L_S^{(d-1)} }{2G} \left[\frac{\delta^{-\mathfrak{B}_0}}{\mathfrak{B}_0}-\frac{(l/2)^{-\mathfrak{B}_0 }G_0^{\mathfrak{B}_0}G_0}{\mathfrak{B}_0}+n\kappa^{(n-1)} \lambda_{R^n}\left[\frac{\delta^{-\mathfrak{B}_1}}{\mathfrak{B}_1}-\frac{(l/2)^{-\mathfrak{B}_1 }G_0^{\mathfrak{B}_1}G_1}{\mathfrak{B}_1} \right] \right]+\mathcal{O}(\lambda_{R^n}^2) \, ,
\ee
where $\mathfrak{B}_1$ is now given by
\be
\mathfrak{B}_1=\frac{2(n-1)\theta}{d}+d-\theta-1\, .
\ee
$G_0$ and $G_1$ are again given by (\ref{gs}) taking the new value of $\mathfrak{B}_1$. As we can see, (\ref{srn}) includes the $\mathcal{O}(\lambda_{R^n})$ correction to the universal term as well as a divergence of order $\mathfrak{B}_1$. This is always subleading with respect to $\mathfrak{B}_0$ and, interestingly, it becomes logarithmic when
\be \label{thetalog}
\theta=\frac{d(d-1)}{d-2(n-1)}\, ,
\ee
provided that $2(n-1)>d$. This value of $\theta$ resembles the $\theta=d-1$ famous result for which a logarithmic divergence is found in the HEE for Einstein gravity ($n=1$), as we will review in a moment. However, this new set of divergences is found for $\theta<0$, whereas the other occurs with $\theta=d-1\geq 0$. Obviously, when $n=2$, the only possibility is $d=1$, which makes $\theta=0$ and reduces to the AdS$_3$ case already studied at the beginning of the section. For $n>2$, however, the situation is much richer, and we find a plethora of new logarithmic divergences (see Figure \ref{logs1}).

When (\ref{thetalog}) is satisfied and $2(n-1)>d$, the HEE expression becomes
\be \label{srnlog}
S_{R^n}=\frac{L ^d L_S^{(d-1)} }{2G} \left[\frac{\delta^{-\mathfrak{B}_0}}{\mathfrak{B}_0}-\frac{(l/2)^{-\mathfrak{B}_0 }G_0^{\mathfrak{B}_0}G_0}{\mathfrak{B}_0}+n\kappa^{(n-1)} \lambda_{R^n}\left[\log\left[\frac{l}{\delta} \right]+c_{R^n}\right] \right]\, ,
\ee
where now
\be
\mathfrak{B}_0=\frac{2(n-1)(d-1)}{2(n-1)-d}\, ,
\ee
and $c_{R^n}$ is a constant correcting the universal term. Therefore, we see that starting from curvature-cubed gravities, introducing higher-order terms in the gravitational action allows one to find new logarithmic contributions to the HEE for hvLf geometries. In both (\ref{srn}) and (\ref{srnlog}) we find a leading divergence whose coefficient scales with the area of the boundary of our entangling region. However, while in (\ref{srn}) the coefficient of the subleading term is also proportional to $\partial A$, in (\ref{srnlog}) we find a different scaling, provided there appears a factor which depends logarithmically on the width of the stripe $l$.

If our guess is right, (\ref{srn}) (and (\ref{srnlog}) when it applies) would be the right expression (swapping $\kappa$, $\lambda_{R^n}$ and so on for the corresponding parameters) for the HEE of a strip in the boundary of a hvLf geometry with $\theta\leq 0$ for any higher-order gravity of $n$-th order in the Riemann tensor.

\section{HEE for hvLf geometries in higher-curvature gravities II: $0<\theta<d$}\label{secr3}
In this section we turn to the case of $0<\theta<d$, corresponding to hvLf metrics whose curvature invariants diverge in the UV (as $r\rightarrow 0$). In order to do so, we follow the steps of \cite{Ogawa:2011bz} and consider these hvLf metrics to be completed asymptotically by an AdS geometry\footnote{See \cite{Dong:2012se} for a different approach, analogous to the one we follow in the previous section.}. Hence, we will assume them to hold only above certain scale $r_F$.

Again, HEE for this class of hvLf spacetimes was studied for Einstein gravity, e.g., in \cite{Ogawa:2011bz} and \cite{Dong:2012se}. In order to be consistent with the conventions in \cite{Ogawa:2011bz}, whose results we plan to generalize here, let us make a change of coordinates in (\ref{hvLf})
\be
r= R^{\frac{d}{(d-\theta)}}\, ,
\ee
and let us relabel $R\rightarrow r$ so there is no confusion between the radial coordinate and the Ricci scalar. Our hvLf geometries read now
\be \label{hvLfz}
ds^2=\frac{L^2}{r^2}\left[-\frac{dt^2}{r^{\frac{2d(z-1)}{d-\theta}}}+r^{\frac{2\theta}{d-\theta}}dr^2+d\vec{x}^2_{(d)} \right]\, .
\ee
The idea is to start with a metric of the form
\be \label{fg}
ds^2=\frac{L^2}{r^2}\left[-f(r)dt^2+g(r)dr^2+d\vec{x}^2_{(d)} \right]\, ,
\ee
and require it to be asymptotically AdS$_{d+2}$ while assuming it to posses some intermediate hvLf-like behaviour
\begin{eqnarray} \label{condd}
g(r)&\simeq&\left[\frac{r}{r_F}\right]^{\frac{2\theta}{d-\theta}}\,,\,\, (r>>r_F)\, ,\\ \notag
g(r)& \simeq& 1 \,,\,\, (r<<r_F)\, ,\\ \notag \\ \notag
f(r)&\simeq&\left[\frac{r}{r_F}\right]^{\frac{2d(1-z)}{d-\theta}}\,,\,\, (r>>r_F)\, ,\\ \notag
f(r)& \simeq& 1 \,,\,\, (r<<r_F)\, .
\end{eqnarray}
Now, if we parametrize the entangling surface as $x_1=F(r)$, computing the induced metric to obtain the area-functional is straightforward, and the result reads \cite{Ogawa:2011bz}
\be
S_{EG}=\frac{L^d L_S^{d-1}}{2G}\int^{r_*}_{\delta} \frac{dr}{r^d}\sqrt{g(r)+\dot{F}(r)^2}\, .
\ee
$r_*$ is the turning point now, where $\dot{F}(r)$ diverges. For this functional there is a first integral given by
\be
\dot{F}=\frac{r^d}{r_*^d}\sqrt{\frac{g(r)}{1-r^{2d}/r_*^{2d}}}\, ,
\ee
so in the end we find
\be \label{segtd}
S_{EG}=\frac{L^d L_S^{d-1}}{2G}\int^{r_*}_{\delta} \frac{dr}{r^d}\sqrt{\frac{g(r)}{1-r^{2d}/r_*^{2d}}}\, .
\ee
The turning point is related to the strip width through
\be \label{tp}
\frac{l}{2}=\int^{r_*}_0 dr \frac{r^d}{r_*^d}\sqrt{\frac{g(r)}{1-r^{2d}/r_*^{2d}}}\, .
\ee
In order to compute these integrals, we need to specify what the exact functional form of $g(r)$ is. However, we can simplify the issue by assuming the entangling surface to probe deep into the IR, so $r_*>>r_F$ \cite{Ogawa:2011bz}. In such a case, (\ref{segtd}) and (\ref{tp}) can be estimated making use of (\ref{condd}), and the result is \cite{Ogawa:2011bz}
\be \label{segoc}
S_{EG}=\frac{L^d L_S^{d-1}}{2G}\left[\frac{\delta^{-(d-1)}}{(d-1)}+\, \frac{c}{r_F^{d-1}}\frac{l^{-\mathfrak{B}_0}}{r_F^{-\mathfrak{B}_0}}+... \right]\, ,
\ee
where $c$ is a numerical constant and the dots refer to subleading contributions which we are neglecting in the limit $r_*>>r_F$. Therefore, we find an area-law term, plus a term which depends on the intermediate scale $r_F$. When $\theta=d-1$, (\ref{segoc}) produces a logarithmic dependence on $r_F$ \cite{Ogawa:2011bz},
\be \label{segoc3}
S_{EG}=\frac{L^d L_S^{d-1}}{2G}\left[\frac{\delta^{-(d-1)}}{(d-1)}+\frac{c}{r_F^{d-1}}\, \log \left[\frac{l}{r_F} \right]+... \right]\, .
\ee
This expression resembles the EE expression expected for a QFT with a Fermi surface \cite{Wolf:2006zzb,arxiv-0908.1724}
\be
S=\alpha \frac{L_S^{d-1}}{\delta^{d-1}}+\beta L_S^{d-1}k_F^{d-1}\log(l k_F)+...\, ,
\ee
being $k_F$ de Fermi momentum and $\alpha$, $\beta$ numerical positive constants. We see that the parameter $r_F$ can be thus interpreted as the Fermi surface scale $r_F\sim k_F^{-1}$.

In order to study the effect of higher-curvature gravities we should repeat the analysis of section \ref{secr2} and start considering curvature-squared gravities one by one. However, taking into account that our approach relies on approximating the spacetime geometry by two different metrics, namely AdS in the UV and hvLf above some scale $r_F$ without specificating its exact form, the calculations for the Gauss-Bonnet and Ricci$^2$ terms become rather filthy and obscure the main goal of this section, which is nothing but studying the kind of terms that one should expect from general higher-order gravities. Therefore, let us stick to $R^n$ gravity, for which we can find the surface extremizing the HEE functional for the general metric (\ref{fg}) and make a treatment as rigorous as the one performed in \cite{Ogawa:2011bz} for Einstein gravity. Following previous steps we find the expression for the HEE functional to be
\be \label{segtdrn}
S_{R^n}=\frac{L^d L_S^{d-1}}{2G}\int^{r_*}_{\delta} \frac{dr}{r^d}T(r)\sqrt{\frac{g(r)}{1-\frac{T(r_*)^2}{T(r)^2}\frac{r^{2d}}{r_*^{2d}}}}\, ,
\ee
where
\be
T(x)\equiv \left[1+n \lambda_{R^n}\tilde{L}^{2(n-1)} R^{(n-1)}(x)\right]\, ,
\ee
with the turning point being related to $l/2$ by
\be \label{tprn}
\frac{l}{2}=\int^{r_*}_0 dr \frac{r^d}{r_*^d}T(r)\sqrt{\frac{g(r)}{1-\frac{T(r_*)^2}{T(r)^2}\frac{r^{2d}}{r_*^{2d}}}}\, .
\ee
It is a tedious but otherwise straightforward calculation to perform the previous on-shell integral and rewrite it in terms of $l$ at order $\mathcal{O}(\lambda_{R^n})$\footnote{It is interesting to note that expanding in powers of $\lambda_{R^n}$ and neglecting higher order contributions is right in this case because the term which goes with the coupling in $T(r)$ scales as $\sim 1/r^{2\theta(n-1)/d}$, with a positive exponent for $\theta>0$, so when we evaluate the integral at $r\rightarrow r_*>>r_F$, the term involving $\lambda_{R^n}$ is small, and the expansion makes sense.}. The final result is
\be \label{segoc4}
S_{R^n}=\frac{L^d L_S^{d-1}}{2G}\left[\frac{\delta^{-(d-1)}}{(d-1)}(1+\lambda_{R^n} c_0)+ \frac{c}{r_F^{d-1}}\frac{l^{-\mathfrak{B}_0}}{r_F^{-\mathfrak{B}_0}}+ \frac{c_1\lambda_{R^n}}{r_F^{d-1}}\frac{l^{-\mathfrak{B}_1}}{r_F^{-\mathfrak{B}_1}}+\mathcal{O}(\lambda_{R^n}^2) \right]\, ,
\ee
where, just as in the $\theta \leq 0$ case
\begin{eqnarray}
\mathfrak{B}_0&\equiv& d-\theta-1\, , \\
\mathfrak{B}_1&\equiv&\mathfrak{B}_0+\frac{2\theta(n-1)}{d}\, ,
\end{eqnarray}
and $c_0$, $c_1$ are numerical constants. As we can see, the kind of terms appearing here resembles those found for $\theta \leq 0$ geometries.  In particular, the term with the power $\mathfrak{B}_1$ produces a logarithmic term when
\be \label{thetalogb}
\theta=\frac{d(d-1)}{d-2(n-1)}\, ,
\ee
as long as $d>2(n-1)$ and $\theta<d$. This seems to generalize the case $\theta=d-1$ to $R^n$ gravities for positive values of the hyperscaling violation exponent. However, $\theta<d$ imposes the following constraint on the order of the gravitational theory admitting such a term
\be
3-2n>0\, ,
\ee
which of course is only satisfied for $n=1$. This reduces to the well-known case of Einstein gravity corresponding to $\theta=d-1$. Therefore, as opposed to the $\theta \leq 0$ case, we do not find additional logarithmic terms in this case for any higher-curvature gravity. Nevertheless, it is not clear that $\mathfrak{B}_1$ is the only new contribution susceptible of arising in this case for general $n$th-order gravities. Further study in this direction would be desirable.


\section{Discussion and perspectives}\label{conc}
In this paper we have considered the effects of higher-order gravity Lagrangians on the HEE expression for geometries with hyperscaling violation. Although the cut-off dependence of the HEE  In section \ref{secr2} we have argued that for $\theta \leq 0$, in order to extract the structure of terms for general higher-curvature gravities, it suffices to evaluate the corresponding on-shell functionals on the extremal area surface, without having to obtain the new surfaces extremizing those functionals, something that would be nevertheless necessary for obtaining the right corrected constant terms. This argument is explicitly illustrated for $R^2$ gravity, for which we can actually extremize the new functional and find the first-order correction to the universal term of the HEE. Our results show that for a general curvature-squared gravity, in addition to the Einstein gravity divergence ($\delta^{-\mathfrak{B}_0}$, with $\mathfrak{B}_0=d-\theta-1$), there appears a single new one, at order $\mathcal{O}(\lambda)$ in the gravitational coupling of the form $\delta^{-\mathfrak{B}_1}$, with $\mathfrak{B}_1=2\theta/d+d-\theta-1$.

The fact that, in spite of the different structure of the corresponding HEE functionals for $R^2$ (\ref{sr2}), Gauss-Bonnet (\ref{sr3}) and Ricci$^2$ (\ref{sr4}) gravities, we find only one divergence of the same order in all cases led us to conjecture that this result extends to arbitrary $n$th-order gravities, so the divergent term found for $R^n$, $\mathfrak{B}_1=2(n-1)\theta/d+d-\theta-1$ , would be the only one appearing for any other theory of that order in curvature when $\theta \leq 0$. It might be that the result does not extend to $n\geq 3$ and that new divergent terms appear when those $n$th-order Lagrangians differ from the simple $R^n$ case. Even if that were the case, that would imply that we are forgetting new contributions, not that $\mathfrak{B}_1$ gets substituted by them. Indeed, the on-shell evaluation of the Wald-like term \cite{Dong:2013qoa}
\be
\frac{\partial \mathcal{L}}{\partial R_{\mu\nu\rho\sigma}}\epsilon_{\mu\nu}\epsilon_{\rho\sigma}\, ,\, \, \text{with}\, \, \epsilon_{\mu\nu}=n^{(\hat{a})}_{\mu}n^{(\hat{b})}_{\nu}\epsilon_{(\hat{a})(\hat{b})}\, ,
\ee
will always contain at least one term scaling with the $(n-1)$th power of the Ricci scalar, which is precisely the one giving rise to $\mathfrak{B}_1$. Therefore, $\mathfrak{B}_1$ will always be there for $n$th-order gravities, although in some cases it might be followed by other divergences appearing for $n\geq 3$.

We have observed that the behaviour arising from Einstein gravity gets corrected for higher-order gravities (at least) by the addition of a new divergent term in which the cut-off scales with a different power, depending on $\theta$, but which is also proportional to the area of the entangling region boundary. Area-law usually tells us about local correlations amongst UV degrees of freedom in the boundary theory. Our findings seem to be suggesting that such correlations get significantly modified when the higher-order couplings are turned on, something which happens to be distinctive of general hvLf geometries with respect to the cases of AdS or Lifshitz without hyperscaling violation, for which the structure of divergences remains unchanged ($\theta=0$ and so $\mathfrak{B}_0=\mathfrak{B}_1=d-1$) and the only difference produced by the inclusion of such terms is a shift on the coefficient in front of $\delta^{-(d-1)}$ (see (\ref{eer5}) \cite{Solodukhin:2009sk}). Nevertheless, it is important to note that, as explained in the introduction, hvLf backgrounds with $\theta \neq 0$ generically suffer from a linearly divergent dilaton in the UV. This obscures the interpretation of the structure of divergences found in the HEE expression in terms of the degrees of freedom of the dual theory (which, to the best of our knowledge, is not known at present for general hvLf backgrounds). The situation is similar to that found for non-conformal branes, where the dual theory is known to be SYM (with $d\neq 4$). In that case, the dilaton, which is related to the YM coupling, also runs in the UV, which means that the theory is either asymptotically free or it needs a UV completion (depending on the dimension). In order to determine what the case is, one needs the exact relation between the dilaton and the coupling. When the YM coupling blows up in the UV, supergravity is not a valid description and S-duality needs to be used. For hvLf metrics, however, the dual theory is not known and the approach taken in the literature is more phenomenological/engineering-like since the supergravity result is taken to \emph{define} what is meant by the dual theory\footnote{We thank again Robert C. Myers and Ioannis Papadimitriou for the explanations appearing in this paragraph.}. Either way, comparing the results found in sections \ref{secr2} and \ref{secr3}, we see that, regardless of the approach we take in computing HEE for these geometries, to wit: either assuming them to be \emph{valid descriptions} in the UV (as in \cite{Dong:2012se}), or considering some AdS completion (as in \cite{Ogawa:2011bz}), we find that the structure of the result does not change, and the novelty is always related to the appearance of a new term $\Lambda^{-\mathfrak{B}_1}$, being $\Lambda$ the scale at which the hvLf geometry is reliable.

Coming back to our results, as we saw, the new term found becomes logarithmic when $d<2(n-1)$ for hvLf geometries with
\be \label{thetaloggg}
\theta=\frac{d(d-1)}{d-2(n-1)}\, ,
\ee
which extends the famous result of $\theta=d-1$ valid for Einstein gravity to negative values of $\theta$. For Einstein gravity ($n=1$) $\mathfrak{B}_0=\mathfrak{B}_1$ and this becomes the leading divergence, whereas in the rest of cases ($n> 1$) we have an area-law-like term with the cut-off scaling as $\delta^{-\mathfrak{B}_0}$ plus the subleading logarithmic term.

Trying to extend this also to the $0<\theta<d$ range, we considered the hvLf geometry to be UV-completed by AdS$_{d+2}$, arising the former above some scale $r_F$ and computed HEE in that case for $R^n$ gravity. We found that $\mathfrak{B}_1$ was the only new contribution again. However, for $0<\theta<d$ we saw that this exponent could not vanish for any $n$ except $n=1$, reducing to the well-known case $\theta=d-1$. In our computation we assume the turning point to probe the IR region, $r_*>>r_F$, in order to be able to approximate the on-shell integrals. It could be that an exact calculation making also use of an exact geometry interpolating between hvLf and AdS in the UV such as the one proposed in \cite{Ogawa:2011bz} gives rise to additional contributions to the HEE when embedded in higher-curvature gravities (and possibly including new logarithmic terms in some cases). Clarifying this possibility and, in general, proving (or refuting) our conjecture on the presence of $\mathfrak{B}_1$ as the only new divergence for general gravities would be interesting. Of course, this looks like a hard task at present.

As we have seen, the fact that all contributions coming from higher-curvature terms are subleading with respect to the Einstein gravity ones forbids these to produce violations of the area law, although we have shown that in certain cases they would yield universal terms which contain factors scaling logarithmically with the stripe width. Therefore, according to our results, only in the exotic case in which the considered gravitational theories did not include the Einstein gravity term could the HEE exhibit new violations of the area law.

\begin{figure}[h]\centering
   \includegraphics[scale=1]{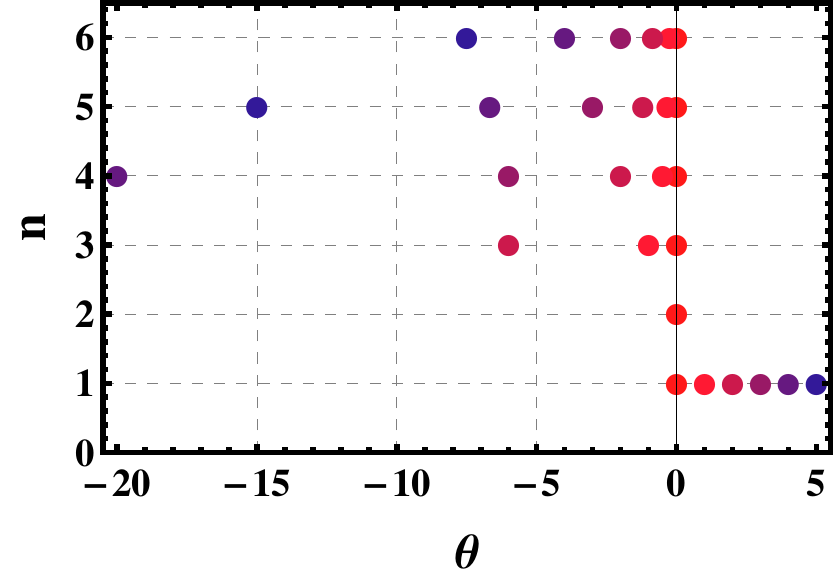}
\caption{Values of $n$ and $\theta$ for which $R^n$ gravities produce logarithmic divergences for different values of $d$. Orange dots correspond to $d=1$ and those in blue to $d=6$.}
\label{logs2}
\end{figure}

In Figure \ref{logs2} we show the values of $n$ and $\theta$ for which $R^n$ (and general $n$th-order gravities) introduce logarithmic terms for different values of $d$. The points on the horizontal line $n=1$ as well as those on the axis $\theta=0$ correspond, respectively, to the cases already known in the literature, namely: hvLf with $\theta=d-1$ and AdS$_3$, whereas those in the quadrant $n>0$, $\theta<0$ are the new ones (extending infinitely for larger values of $n$ and $-\theta$).

Finally, the results obtained here should be extendable to other entangling regions different from the strip, such as cylinders, $m$-spheres and, ideally, arbitrary entangling regions. In principle, we expect subleading divergences to appear when more \emph{complicated} entangling surfaces are considered. These would be produced by geometric integrals along the entangling surface (see \cite{Ryu:2006ef} for an account of this for pure AdS$_{d+2}$). It would be of most interest to investigate how these divergences get modified in hvLf backgrounds. For n-spheres, for example, this has not been accomplished yet (to the best of our knowledge); not even in the simplest case of Einstein gravity.

\acknowledgments
The authors thank Robert C. Myers and Tom\'as Ort\'in for very useful comments and corrections on the manuscript and Ioannis Papadimitriou for illuminating explanations on the UV fate of hvLf geometries. PB wishes to acknowledge Robert C. Myers his invaluable guidance into the holographic world. The work of PB has been supported by the JAE-predoc grant JAEPre 2011 00452. PFR is supported by the grant \emph{Ayuda para contrato predoctoral Severo Ochoa} SVP-2013-067903. The authors are also partially supported by the Spanish Ministry of Science and Education grant FPA2012-35043-C02-01, the \emph{Comunidad de Madrid} grant HEPHACOS S2009ESP-1473, and the Spanish Consolider-Ingenio 2010 program CPAN CSD2007-00042. Research at IFT is supported by the Spanish MINECO's \emph{Centro de Excelencia Severo Ochoa} Programme under grant SEV-2012-0249.

\appendix
\section{HEE for a doubly-Wick-rotated hvLf geometry}\label{secr4}




In this appendix we study HEE for a class of geometries for which the anisotropic scaling occurs along one of the spatial dimensions instead of time \cite{deBoer:2011wk,Alishahiha:2012cm}
\be \label{hvdw}
ds^2=L^2 r^{\frac{2\theta}{d}} \left(-\frac{dt^2}{r^{2}}+ \frac{dr^2}{r^{2}}+\frac{d\vec{x}_{\left(d-1\right)}^2}{r^{2}} +\frac{dy^2}{r^{2z}}  \right) \, .
\ee
This can be understood as obtained through a double Wick rotation of the usual hvLf metric (\ref{hvLf}). Indeed we just have to apply the following transformation to it
\be
t \rightarrow iy \, , \, x_d \rightarrow it \, ,
\ee
where $x_d$ stands for the $d$th spatial coordinate. This makes the geometry covariant under the following transformations
\be
y\rightarrow \lambda^z t\, ,\, t \rightarrow \lambda t \, , \,  x_i\rightarrow \lambda x_i\, ,\,  i=1,...,d-1\, .
\ee

HEE in the framework of Einstein gravity has been already studied for this geometry in \cite{deBoer:2011wk,Alishahiha:2012cm}. Here we are going to extend the study to the case of $R^n$ gravity to illustrate how the result changes with respect to the usual hvLf case. The motivation to consider such a perversion is to make the dynamical exponent $z$ appear in the exponents of the divergent terms in the HEE expression. This indeed results in the production of new divergences, which become logarithmic in a certain subset of the parameter space.

 The region at the boundary for which we compute the entanglement entropy is the same as in the rest of the article, with the particularity that now we have anisotropic spatial scaling. We consider the strip to extend infinitely (up to the IR cut-off $L_S\rightarrow \infty$) along the special scaling coordinate, so $s=\left\{(t_E,r,x_1,x_2,...,x_{d-1},y)\text{ s.t., }t_E=0,\right.$ $\, x_{d-1}\in[-l/2,l/2],$ $\left. \, x_{1,...,d-2}\in (-L_S/2,+L_S/2),\right.$ $\left. \, y\in (-L_S/2,+L_S/2)\right\}$. The procedure used here is the same as that of section (\ref{secr2}), so we will skip redundant discussions.

The HEE functional is 
\begin{eqnarray} \hspace{-1cm}
S_{R^n}=\frac{1}{4G} \int_{m} d^2y \sqrt{g_{m}}\left[1+n \lambda_{R^n} \tilde{L}^{2(n-1)} R^{n-1}\right]\, .
\end{eqnarray}
The Ricci scalar for (\ref{hvdw}) is the same as that for (\ref{hvLf}), that is, $R=\kappa r^{-2\theta/d}/\tilde{L^2}$. We can parametrize the entangling surface $m$ as $x_{d-1}=h(r)$, so that the metric induced in such surface is
\begin{eqnarray}
ds_m^2=L^2r^{\frac{2\theta}{d}} \left[ \frac{dy^2}{r^{2z}} + \left( 1+\dot{h}^2 \right) \frac{dr^2}{r^2}+\frac{d \vec{x}_{d-2}^2}{r^2} \right] \, ,
\end{eqnarray}
The expression for the entanglement entropy becomes
\be
S_{R^n}=\frac{L^d L_S^{(d-1)}}{2G} \int_{\delta}^{r_*} dr \sqrt{1+\dot{h}^2} f(r) r^{(\theta-d-z+1)} \, , \, \text{with} \, f(x) \equiv \left[1+n\kappa^{(n-1)} \lambda_{R^n} x^{-2\theta(n-1)/d}\right] \, ,
\ee
$r_*$ being the turning point of the surface, where $\dot{h} \vert_{r_*}=\infty$. The functional has a first integral associated to $h$, so we can express $\dot{h}$ in terms of $h$. By doing so and after some rearrangement we find
\be
S_{R^n}=\frac{L^d L_S^{(d-1)} r_*^{\theta-d-z+2}}{2G} \int_{\delta/r_*}^{1} du \frac{u^{(\theta-d-z+1)} f(u h_*)}{\sqrt{1-u^{2(d-\theta+z-1)}\frac{f(r_*)^2}{f(u r_*)^2}}} \, .
\ee
We need $d-\theta+z-1>0$ for the perturbative analysis to be consistent. Under this condition the expression looks exactly like the one in section \ref{secr2} after promoting $(d-\theta) \rightarrow (d-\theta+z-1)$. This implies the following result for the HEE
\be
S_{R^n}=\frac{L ^d L_S^{(d-1)} }{2G} \left\{\frac{\delta^{-\mathfrak{B}_0}}{\mathfrak{B}_0}-\frac{(l/2)^{-\mathfrak{B}_0 }G_0^{\mathfrak{B}_0}G_0}{\mathfrak{B}_0}+n\kappa^{(n-1)} \lambda_{R^n}\left[\frac{\delta^{-\mathfrak{B}_1}}{\mathfrak{B}_1}-\frac{(l/2)^{-\mathfrak{B}_1 }G_0^{\mathfrak{B}_1}G_1}{\mathfrak{B}_1} \right] \right\}+\mathcal{O}(\lambda_{R^n}^2) \, ,
\ee
with
\begin{eqnarray}
\mathfrak{B}_0&\equiv& d-\theta+z-2\, , \\
\mathfrak{B}_1&\equiv&\mathfrak{B}_0+\frac{2\theta(n-1)}{d}\, ,
\end{eqnarray}
\be
G_0\equiv\frac{\sqrt{\pi}\Gamma\left(\frac{\mathfrak{B}_0+2}{2(\mathfrak{B}_0+1)}\right)}{\Gamma\left(\frac{1}{2(\mathfrak{B}_0+1)}\right)} \, , \,\, \,G_1\equiv\frac{ \sqrt{\pi} \Gamma\left(\frac{2
+2\mathfrak{B}_0-\mathfrak{B}_1}{2(\mathfrak{B}_0+1)}\right)     }{ \Gamma\left(\frac{1+\mathfrak{B}_0-\mathfrak{B}_1}{2(\mathfrak{B}_0+1)}\right)}\, .
\ee
The divergence with $\mathfrak{B}_1$ becomes logarithmic when
\be \label{logdw}
\theta=\frac{d(d+z-2)}{d-2(n-1)} \, ,
\ee
which gives a broad range of possibilities. However, we still need to take into account the NEC, which are different with respect to those for the standard hvLf case. For Einstein gravity, this is computed as $G_{\mu\nu}N^{\mu}N^{\nu}\geq 0$, $N^{\mu}$ being appropriate null vectors and $G_{\mu\nu}$ the Einstein tensor. For higher-curvature gravities, we will find additional conditions involving the couplings of the theory, which we assume to be susceptible of being satisfied by tuning those. For this metric a convenient null vector is
\be
N^r=\frac{s_r}{L}r^{1-\theta/d} \, , \, N^i=\frac{s_i}{L}r^{1-\theta/d} \, , \, N^y=\frac{s_y}{L}r^{z-\theta/d} \, ,
\ee
\be
N^t=\frac{\sqrt{\sum s_i^2+s_r^2+s_y^2}}{L} r^{1-\theta/d} \, .
\ee
with the $s_{\mu}$ being positive constants. The NEC produces two inequalities
\begin{eqnarray}
d(z-1)z+\theta(d-\theta) &\leq&0 \, , \\
(z-1)(z+d-\theta) &\leq&0 \, .
\end{eqnarray}
After some algebra, one can see that these limit the allowed values of $z$ to lie in the interval
\be \label{necdw}
\frac{1-\sqrt{1+4\theta\frac{\theta-d}{d}}}{2} \leq z \leq 1 \, .
\ee
So for each dimension $d$ and each order in curvature $n$, any metric with $z$ satisfying (\ref{necdw}) will give rise to a logarithmic contribution as long as (\ref{logdw}) is satisfied.

\renewcommand{\leftmark}{\MakeUppercase{Bibliography}}
\phantomsection
\bibliographystyle{JHEP}
\bibliography{References}
\label{biblio}

\end{document}